\documentclass[letterpaper]{jpconf}
\usepackage{graphicx}
\begin{document}
\title{The Phononic Casimir Effect: An Analog Model}

\author{L.H. Ford}
\address{Institute of Cosmology,
Department of Physics and Astronomy \\
         Tufts University, Medford, MA 02155}
\ead{ford@cosmos.phy.tufts.edu}
 
\author{N.F. Svaiter }
\address{Centro Brasiliero de Pesquisas Fisicas CBPF \\
 Rua Dr. Xavier Sigaud 150,
Rio de Janeiro, RJ, 22290 180, Brazil}
\ead{nfuxsvai@cbpf.br}

\begin{abstract}
We discuss the quantization of sound waves in a fluid with a linear dispersion relation
and calculate the quantum density fluctuations of the fluid in several cases. These
include a fluid in its ground state. In this case, we discuss the scattering cross 
section of light by the density fluctuations, and find that in many situations it
is small compared to the thermal fluctuations, but not negligibly small and might be
observable at room temperature. We also consider a fluid in a squeezed state of phonons
and fluids containing boundaries. We suggest that the latter may be a useful analog model
for better understanding boundary effects in quantum field theory. In all cases 
involving boundaries which we consider, the mean squared density fluctuations are 
reduced by the presence of the boundary. This implies a reduction in the light scattering
cross section, which is potentially an observable effect.
\end{abstract}

\section{Introduction}

It is well known that quantized sound waves, whose excitations are phonons, share several
properties with relativistic quantum fields, such as the electromagnetic field. Here we will
be primarily concerned with fluctuations in the phonon ground state. These fluctuations
can share certain features with electromagnetic vacuum fluctuations, including the creation
of Casimir-type forces~\cite{DLP}. The forces due to quantum or stochastic sound fluctuations
have been discussed recently by several authors~\cite{Larraza,Bschorr,SU02,Recati,Lamoreaux}.
In this paper, we will study quantum density fluctuations in a fluid. The fluid in question
could be either a classical or a quantum fluid, but we will consider only the regime where
the dispersion relation is approximately linear. Thus the analog models which we consider
are rather different than those which model the Hawking process~\cite{Unruh,Jacobson,BLV05}.
The topic of hydrodynamic fluctuations has been reviewed, for example, in Refs.~\cite{Foster,ZS06}.
 In Sec.~\ref{sec:nobound}, we will discuss
density fluctuations in a fluid in the phonon ground state without boundaries, and summarize 
work recently reported in
Ref.~\cite{FS08}. It was argued there that light scattering by quantum density fluctuations might be
observable, even in fluids at room temperature. The effects of a squeezed state of phonons 
will be briefly discussed in Sec.~\ref{sec:squeezed}.
In Sec.~\ref{sec:bound}, we turn to the issue
of the effects of boundaries in the fluid, and argue that this can be a useful analog model
for the effects of boundaries on relativistic quantum fields. We also suggest that the
boundary effects, at least in principle, create observable effects in the fluid case. 
We report on the result of several specific calculations for different geometries.
The results will be summarized and discussed in Sec.~\ref{sec:final}.

\section{Phonons and Density Fluctuations in a Fluid} \label{sec:nobound}
\subsection{Quantization and the Density Correlation Function}

 We consider the quantization of sound waves in a fluid with a
linear dispersion relation, $\Omega_q = c_S\, q$, where  $\Omega_q$
is the phonon angular frequency, $q$ is the magnitude of the wave
vector, and $c_S$ is the
speed of sound in the fluid. This should be a good approximation for
wavelengths much longer than the interatomic separation.  
Let $\rho_0$ be the mean mass density
of the fluid. Then the variation in density around this mean value is
represented by a quantum operator, $\hat{\rho}(\mathbf{x},t)$, which
may be expanded in terms of phonon annihilation and creation operators
as~\cite{LL-ST}
\begin{equation}
 \hat{\rho}(\mathbf{x},t) = \sum_{\mathbf{q}} (b_{\mathbf{q}}
 f_{\mathbf{q}} + b^\dagger_{\mathbf{q}}  f^*_{\mathbf{q}}) \,,
\end{equation}
where
\begin{equation}
 f_{\mathbf{q}} = \sqrt{\frac{\hbar \omega \rho_0}{2 V c_S^2}}
\; {\rm e}^{i(\mathbf{q} \cdot \mathbf{x} -\Omega_q\, t)} \,. 
\label{eq:mode_fnt}
\end{equation}
Here $V$ is a quantization volume. The normalization factor in 
Eq.~(\ref{eq:mode_fnt}) can be fixed by requiring that the zero point energy
of each mode be $\frac{1}{2} \hbar \Omega_q$ and using the expression for the 
energy density in a sound wave,
\begin{equation}
U = \frac{c_S^2}{\rho_0}\, {\hat{\rho}}^2\,.
\end{equation}
 In the limit in which $V
\rightarrow \infty$, we may write the density correlation function
as
\begin{equation}
\langle \hat{\rho}(\mathbf{x},t)\, \hat{\rho}(\mathbf{x}',t') \rangle
= \frac {\hbar \rho_0}{16 \pi^3 c_S^2}\, \int d^3q\, \Omega_q\,
 {\rm e}^{i(\mathbf{q} \cdot\Delta \mathbf{x} -\Omega_q\, \Delta t)}\,, 
                        \label{eq:rhorho}
\end{equation}
where $\Delta \mathbf{x} = \mathbf{x} - \mathbf{x}'$ and
$\Delta t = t-t'$. The integral may be evaluated to write the
coordinate space correlation function as
\begin{equation}
\langle \hat{\rho}(\mathbf{x},t)\, \hat{\rho}(\mathbf{x}',t') \rangle
= -\frac{\hbar \rho_0}{2 \pi^2 c_S}\; 
\frac{\Delta \mathbf{x}^2 +3 c_S^2 \Delta t^2}
{(\Delta \mathbf{x}^2 - c_S^2 \Delta t^2)^3}\,.
\end{equation}

This is of the same form as the correlation function for the time
derivative of a massless scalar field in relativistic quantum field
theory, 
$\langle \dot{\varphi}(\mathbf{x},t)\, \dot{\varphi}(\mathbf{x}',t')
\rangle$.
(This analogy has been noted previously in the literature. See, for example, 
Ref.~\cite{FF04}.)
Apart from a factor of $\rho_0$, these two quantities may be obtained
from one another by interchanging the speed of light $c$ and the speed of
sound $c_S$. If $c \rightarrow c_S$, then 
\begin{equation}
\langle \dot{\varphi}(\mathbf{x},t)\, \dot{\varphi}(\mathbf{x}',t')
\rangle \rightarrow \rho_0 \,
\langle \hat{\rho}(\mathbf{x},t)\, \hat{\rho}(\mathbf{x}',t')
\rangle\,.
\end{equation}

In the limit of equal times, the density correlation function becomes
\begin{equation}
\langle \hat{\rho}(\mathbf{x},t)\, \hat{\rho}(\mathbf{x}',t) \rangle
= -\frac{\hbar \rho_0}{2 \pi^2 c_S \, (\Delta \mathbf{x})^4} \,.
\label{eq:rho=t}
\end{equation}
Thus the density fluctuations increase as $|\Delta \mathbf{x}|$
decreases. Of course, the continuum description of the fluid and the
linear dispersion relation both fail as  $|\Delta \mathbf{x}|$
approaches the interatomic separation. Also note the minus sign in
Eq.~(\ref{eq:rho=t}). This implies that density fluctuations at
different locations at equal times are anticorrelated. By contrast,
when  $c_S|\Delta t| > |\Delta \mathbf{x}|$, then
$\langle \hat{\rho}(\mathbf{x},t)\, \hat{\rho}(\mathbf{x}',t) \rangle >0$
and the fluctuations are
positively correlated. This is complete analogy with the situation in
the relativistic theory. Fluctuations inside the lightcone can
propagate causally and tend to be positively correlated. Fluctuations 
in a fluid 
for which $c_S|\Delta t| < |\Delta \mathbf{x}|$ cannot have propagated
from one point to the other, and are anti-correlated. This can be
understood physically because an over density of fluid at one point
in space requires an under density at a nearby point.

\subsection{Light Scattering by Density Fluctuations}

In Ref.~\cite{FS08}, the cross section for the scattering of light by the zero point
 density fluctuations is computed for the case that the incident light
angular frequency is large compared to the typical phonon frequency.
 The result is
 \begin{equation}
\left(\frac{d\sigma}{d\Omega}\right)_{ZP} = \sqrt{2(1-\cos \theta)}\;
\frac{\hbar \omega^5\, {\cal V}\, \eta^4}{32
  \pi^2 c^5 c_S\rho_0}\;
(\mathbf{\hat{e}}_{{\mathbf k},\lambda} \cdot 
\mathbf{\hat{e}}_{{\mathbf k}',\lambda'})^2    \,, 
\label{eq:cross_ZP2}
\end{equation}
where $\theta$ is the scattering angle, ${\cal V}$ is the scattering
volume, and $\eta$ is the mean index of refraction of the fluid.
 The $\omega^5$ dependence
of the scattering cross section can be viewed as the product of the
$\omega^4$ dependence of Rayleigh-Brillouin scattering and one power
of $\Omega_q$, and hence of $\omega$, coming from the spectrum of
zero point fluctuations in the fluid. The factor of $\eta^4$ represents
the influence of the fluid on light propagation before and after the
scattering process, and arises as a product of a factor of $\eta$ in the
incident flux and a factor of $\eta^3$ in the density of final states~\cite{FS08}.  
Because light travels through 
the fluid at speeds much greater than the sound speed, light
scattering  reveals a nearly static distribution of density
fluctuations. Thus we can regard Eq.~(\ref{eq:cross_ZP2}) as a
probe of the fluctuations described by Eq.~(\ref{eq:rho=t}).
The scattering by  zero point fluctuations is inelastic, with
the creation of a phonon. Thus, the scattering described by
Eq.~(\ref{eq:cross_ZP2}) is strictly Brillouin rather than Rayleigh scattering.

This scattering by zero point density fluctuations should be compared to
the effects of thermal density fluctuations. The ratio of the zero point to the
thermal scattering can be expressed as
 \begin{equation}
R \equiv \frac{(d\sigma/d\Omega)_{ZP}}{(d\sigma/d\Omega)_{TB}} =
 \sqrt{2(1-\cos \theta)}\,
\left(\frac{\hbar \omega}{2 k_B T}\right)\,
\left(\frac{c_S}{c}\right)\, \eta^4\, \left[
 \rho_0\, \left(\frac{\partial \epsilon}{\partial
  \rho_0}\right)_S \right]^{-2}\,.
\end{equation}   
The index of refraction, $\eta$, and the quantity 
$ \rho_0\, \left({\partial \epsilon}/{\partial  \rho_0}\right)_S$,
which involves a derivative of the fluid dielectric function with respect
to density at constant entropy
are both of order unity. Hence $R$ is primarily determined by the ratio of the
photon energy to the thermal energy, and the ratio of the speed of
sound to the speed of light.

As an example, consider the case of water at room temperature and
violet light with a wavelength of $\lambda = 350nm$. In this case,
we have $c_S = 1480 m/s$ and $\eta =1.4$~\cite{CRC}. In addition, 
$ \rho_0\, \left({\partial \epsilon}/{\partial  \rho_0}\right)_S =
0.79$ ~\cite {CG66}. For back scattering, $\cos \theta = -1$,
this leads to $R \approx 0.005$. Consequently, about $0.5\%$
of the Stokes line is due to zero point motion effects. Although this
is a small fraction, it may be detectable, and will increase at lower
temperatures and shorter wavelengths.

\section{Squeezed States of Phonons}\label{sec:squeezed}

Now we consider the case where the phonon field is not in the vacuum state,
but rather a squeezed state. The squeezed states are a two complex parameter family
of states, but we will focus attention on the case of the squeezed vacuum 
states $|\zeta\rangle$,
labeled by a single complex squeeze parameter
\begin{equation}
\zeta = r\,{\rm e}^{i\delta} \,.
\end{equation}
This set of states is of special interest because they are the states generated by
quantum particle creation processes, and they can exhibit local negative energy
densities. Consider the shift in the mean squared density fluctuations between
the given state and the vacuum
\begin{equation}
\langle \hat{\rho}^2 \rangle_R = \langle \zeta|\hat{\rho}^2|\zeta \rangle
- \langle 0|\hat{\rho}^2|0 \rangle \,,
\end{equation}
the ``renormalized'' mean squared density fluctuation. The result for this quantity
in a single mode squeezed vacuum state for a plane wave in the $z$-direction is
\begin{equation}
\langle \hat{\rho}^2 \rangle_R = 
\frac{ \hbar\, \omega\,\rho_0}{c_S^2 V}\, \sinh r \;\left\{ \sinh r
-\cosh r\, \cos[2(kz-\omega t) + \delta] \right\}\,.
\end{equation}
Note that this quantity can be either positive or negative, but its time or space average
is positive. The suppression of the local density fluctuations in a squeezed state
is analogous to the creation of negative energy densities for a massless, relativistic
field.

\section{Boundaries}\label{sec:bound}

If we introduce an impenetrable boundary into the fluid, the phonon field will
satisfy Neumann boundary conditions
\begin{equation}
{\bf \hat{n} \cdot \nabla}  \delta \rho =0
\end{equation}
as a consequence of the impenetrability. Thus there will be a Casimir force on the
boundaries which is analogous to the Casimir force produced by electromagnetic
vacuum effects. For example, consider two parallel plates, which will experience
an attractive force per unit area of
\begin{equation}
\frac{F}{A} = \frac{\hbar \, c_S\, \pi^2}{480\, a^4}\,,
\end{equation}
which is smaller than the electromagnetic case for perfect plates by a factor of
${c_S}/(2c)$, and is thus quite small in any realistic situation. 

Henceforth, we consider the local effect of boundaries on mean squared density
 fluctuations, and now define $\langle \hat{\rho}^2 \rangle_R$ to be the change due to
the presence of the boundary. This quantity is of interest both as an analog model for 
the effects of boundaries in quantum field theory, and in its own right. The shifts in
density fluctuations are at least in principle observable in light scattering 
experiments.

Our interest in the phononic analog model is inspired by the fact that
the study of boundary effects
in quantum field theory is an active area of research, and has given rise to
some recent controversies in the literature~\cite{Jaffe,MCW06}.
 One question is the nature of the physical cutoff
which prevents singularities at the boundary. 
An example of the subtleties is afforded by the mean
squared electric and magnetic fields near a dielectric interface. When the
material is a perfect conductor, these quantities are proportional to
$z^{-4}$, where $z$ is the distance to the interface. Specifically, in 
Lorentz-Heaviside units their asymptotic forms are
\begin{equation}
\left\langle E^{2}\right\rangle \sim \frac{3}{16\pi ^{2}}\frac{1}{z^{4}}
\end{equation}
and
\begin{equation}
\left\langle B^{2}\right\rangle \sim -\frac{3}{16\pi ^{2}}\frac{1}{z^{4}}
\end{equation}
 One might expect
that a realistic frequency dependent dielectric function would remove
this singularity, but this is not the case. Instead one finds~\cite{SF05} that
\begin{equation}
\left\langle E^{2}\right\rangle \sim \frac{\sqrt{2}\omega _{p}}{32\pi } 
\frac{1}{z^{3}}
\end{equation}
and
\begin{equation}
\left\langle B^{2}\right\rangle \sim -\frac{5\omega _{p}^{2}}{96\pi }
\frac{1}{z^{2}} \,,
\end{equation}
where $\omega_p$ is the plasma frequency of the material.
Thus some physical cutoff other than dispersion is required. For realistic
materials, it is likely to be surface roughness, but fluctuations in the
position of the boundary can also serve as a cutoff, as we showed
several years ago~\cite{FS98}.  In a fluid, there is always a physical cutoff at the
interatomic separation, but in a given situation other physical
cutoffs may be dominant. For the present, we will report the results of 
calculations of $\langle \hat{\rho}^2 \rangle_R$ for several different geometries,
and leave a more detailed treatment of physical cutoffs for later work.

In the remainder of this section, we will quote several results for 
$\langle \hat{\rho}^2 \rangle_R$ in different geometries. The details of the calculations
involved will be presented in a later paper~\cite{FS08b}.

\subsection{One or Two Parallel Plane Boundaries}

In both of these case, the renormalized density two-point function may be constructed
by the method of images. For the case of a single plane, one finds
\begin{equation}
\langle \hat{\rho}^2 \rangle_R = -\frac{\hbar\, \rho_0\, c_S}{32 \pi^2\, z^4} < 0
\end{equation}
where $z$ is the distance to the boundary. For the case of two parallel planes, the
result is
\begin{equation}
\langle   \hat{\rho}^2 \rangle_R = -\frac{ \hbar\, \rho_0\, c_S}{96  a^4}\;
\left[ \frac{1}{15} +
\frac{3 - 2 \sin^2(\pi z/a)}{ \sin^4(\pi z/a)} \right]\,,
\end{equation}
where $a$ is the separation of the two planes, and $z$ is the distance to one boundary.  
Note that both of these expression are negative everywhere. In the absence of a physical
cutoff, both of these expressions diverge as $z^{-4}$ near the boundaries, just as 
do the squared electric and magnetic fields near a perfectly reflecting plane.

\subsection{A Three-Dimensional Torus}

Here we consider a rectangular box with periodic boundary conditions in all three
spatial directions, with periodicity lengths $L_1$, $L_2$ and $L_3$. Thus the three-dimensional
space has the topology of $S^1\times S^1 \times S^1$. This is closely related
to the geometry of a waveguide, where the fluctuations of a relativistic scalar field
were discussed by Rodrigues and Svaiter~\cite{RS03}.
 As in the parallel plane case, an image
sum method may be employed, with the result
\begin{equation}
\langle  \hat{\rho}^2 \rangle_R = -\frac{ \hbar\, \rho_0\, c_S}{2 \pi^2}\;
{\sum_{\ell,m,n}}' \frac{1}{(\ell^2 L_1^2 +m^2 L_2^2 +n^2 L_3^2)^2}\,.
\end{equation}
Here the prime on the summation indices denotes that the $\ell=m=n=0$ term is
omitted. Here $\langle \hat{\rho}^2 \rangle_R$ is a negative constant.

\subsection{A Wedge}

Consider two intersecting plane which are at an angle of $\alpha$ with respect to each
other. Now consider a point inside of this wedge which is located at polar coordinates
$(r,\theta)$, where $r$ is the distance to the intersection line and $\theta < \alpha$.
This geometry was treated for the relativistic case by Candelas and Deutsch~\cite{CD79}, 
and we may transcript one of their results [their Eq.~(5.39)] to find
\begin{eqnarray}
\langle  \hat{\rho}^2 \rangle_R &=& 
-\frac{ \hbar\, \rho_0\, c_S}{1440 \pi^2\,r^4\, \sin^4(\pi \theta/\alpha)} 
\nonumber \\ &\times&
\left\{(\pi-\alpha)(\pi+\alpha)  \sin^2(\pi \theta/\alpha) [(\pi^2 +11\alpha^2) 
 \sin^2(\pi \theta/\alpha) -30\pi^2] +45\pi^4 \right\}\,.
\end{eqnarray}
Again, this quantity is negative everywhere. 

\subsection{A Cosmic String}

As is well known, the space surrounding a cosmic string is a conical space
with a deficit angle $\alpha < 2\pi$. Quantum field theory in this conical space
has been discussed by many authors, beginning with Helliwell and Konkowski~\cite{HK86},
and is similar to the wedge problem discussed above. We may follow the procedure in
Ref.~\cite{HK86}.   At a distance $r$ from the apex, we find
\begin{equation}
\langle  \hat{\rho}^2 \rangle_R = 
-\frac{ \hbar\, \rho_0\, c_S}{1440 \pi^2\, \alpha^4\, r^4}\;(2\pi-\alpha)(2\pi+\alpha)
(11\alpha^2 +4\pi^2) \,,
\end{equation}
which is also negative everywhere.

\subsection{Near the Focus of a Parabolic Mirror}

\begin{figure}
\begin{center}
     \scalebox{0.3}{ 
\includegraphics{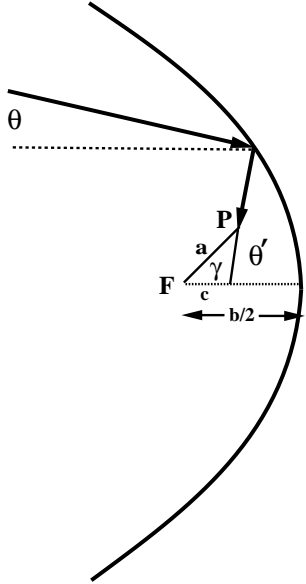}}
\end{center}
\caption{The geometry of rays reflecting from a parabolic mirror is illustrated.
 An incoming ray at an angle of $\theta$ reflects at an angle of
$\theta'$ to reach the point $P$, which is a distance $a$ from the focus $F$, and at an angle
of $\gamma$.}
\label{fig:para}
\end{figure}

The quantization of the electromagnetic field in the presence of a parabolic mirror
was discussed by us in Refs.~\cite{FS00,FS02}, where a geometric optics approximation
was employed to find the mean squared fields near the focus. This treatment lead to the 
result that these quantities are singular at the focus, diverging as an inverse power
of the distance $a$ to the focus. This result holds both for parabolic cylinders and for
parabolas of revolution, and basically arises from the interference term of multiply
reflected rays with nearly the same optical path length. The geometry is illustrated in
Fig.~\ref{fig:para}. An incoming ray at an angle of $\theta$ reflects at an angle of
$\theta'$ to reach the point $P$, which is a distance $a$ from the focus $F$, as illustrated.
The distance from the focus to the mirror itself is $b/2 \gg a$. There can be two values
of $\theta'$, denoted $\alpha$ and $\beta$, corresponding to two reflected rays. The difference
in the optical paths of these two rays is denoted by $\Delta \ell$.

 The detailed expression for
this distance $\Delta \ell$ used in Refs.~\cite{FS00,FS02} is not quite correct, 
as was pointed out to us by Vuletic ~\cite{Vul}. The corrected expression is
\begin{equation}
\Delta \ell = a\,\left[ \cos\gamma (\cos\alpha -cos\beta +\sin^2\alpha -\sin^2\beta)
+\sin\gamma (\sin\alpha -\sin\beta +\sin\beta \cos\beta -\sin\alpha \cos\alpha) \right]\,.
\end{equation}
This is to be used in the expression obtained from geometric optics,
\begin{equation}
\langle E^2 \rangle = {4}{5 \pi^3} \int \frac{d\theta}{(\Delta \ell)^4} \,. \label{eq:E2}
\end{equation}
In this expression, $\alpha=\alpha(\theta)$, and $\beta=\beta(\theta)$, as will be 
discussed in Ref.~\cite{FS08b}.

A detailed discussion of the electromagnetic case will be given elsewhere. Here
we are concerned with $\langle  \hat{\rho}^2 \rangle_R$, which is obtained from
Eq.~(\ref{eq:E2}) by letting $c \rightarrow c_S$ and dividing by $2$, leading to a result
of the form    
\begin{equation}
\langle  \hat{\rho}^2 \rangle_R = 
-\frac{ \hbar\, \rho_0\, c_S\, C}{b\, a^3} <0 \,. \label{eq:near_focus}
\end{equation}
This, and the analogous expressions for $\langle E^2 \rangle$ and $\langle B^2 \rangle$,
which also are proportional to $1/(b\, a^3)$, are striking in that they can be large
when the focus is far from the mirror itself, $b \gg a$. This result is controversial,
and seems to be in conflict with a general result by Fewster and Pfenning~\cite{FP06},
which implies that quantities such as $\langle E^2 \rangle$ or $\langle  \hat{\rho}^2 \rangle_R$
should be proportional to
the inverse fourth power of the distance to the mirror, which is to say 
$\propto b^{-4}$ in this case. On the other hand, there is a simple physical argument
to the contrary in this case, which we find compelling: 
the interference term between multiply reflected rays
is slowly oscillating when $\Delta \ell \propto a$ is small, and should give a 
contribution proportional to an inverse power of $a$, as in Eq.~(\ref{eq:near_focus}). 
In any case, the study of the phononic case provides an additional theoretical,
and potentially experimental, probe to better understand this issue.

\section{Summary}
\label{sec:final}

In this paper, we have considered a classical or quantum fluid with a linear dispersion 
relation as a analog model for quantum fluctuations, and the effects of boundaries on these 
fluctuations. We have argued that the local density fluctuations in a fluid in the phonon
ground state are potentially observable in light scattering experiments. This is of interest
in its own right. We have reported the calculation of the change in these fluctuations due
to squeezed states of phonons and due to the presence of boundaries. Squeezed states can 
produce either positive or negative values for
 $\langle \hat{\rho}^2 \rangle_R$, in a way which is analogous to the
effects of squeezed states on the energy density of a relativistic quantum field. However,
all of the boundary examples which we consider result in a negative
$\langle \hat{\rho}^2 \rangle_R$. If we compare to the already negative result in 
Eq.~(\ref{eq:rho=t}), this seems to represent an increase in magnitude of an already negative
quantity.
 If one were to scatter light from the fluid with a boundary,
this implies a change in the scattering cross section, compared to the result for a 
boundary-free fluid, Eq.~(\ref{eq:cross_ZP2}), at least for short wavelengths. In general,
the calculation of the modified cross section in the presence of a boundary is a complex
task. However, if the incident wavelength is short compared to the scale over which
$\langle \hat{\rho}^2 \rangle_R$ varies, then we might the scattering
cross section to increase in the presence of a boundary. 
A more detailed analysis of this effect is a subject for future study.
It will also be of interest to consider the effects on non-linearity in the phonon
dispersion relation on the questions considered here.

\ack
This work was supported in part by the National Science Foundation under Grant PHY-0555754
 and by Conselho Nacional de Desenvolvimento
Cientifico e Tecnologico do Brasil (CNPq).


\vspace{0.8cm}


\begin{thebibliography}{9}

\bibitem{DLP} I.E. Dzyaloshinskii, E.M. Lifshitz and L.P. Pitaevski 1961
Adv. Phys. {\bf 10}, 165 

\bibitem{Larraza} A. Larraza 1998 Phys. Lett. A {\bf 248}, 151 

\bibitem{Bschorr} O. Bschorr 1999 J. Acoust. Soc. Am. {\bf 106}, 3730 

\bibitem{SU02} E. Sch{\"a}ffer and U. Steiner 2002 Eur. Phys. J. E {\bf 8},
347 

\bibitem{RP05} D.C. Roberts and Y. Pomeau 2005 Phys. Rev. Lett. {\bf 95},
145303 

\bibitem{Recati} A. Recati, J.N. Fuchs, C.S. Pe{\c c}a, and
  W. Zwerger 2005 Phys. Rev. A {\bf 72}, 023616 

\bibitem{Lamoreaux} S.K. Lamoreaux 2008  (Preprint arxiv:0808.4000)

\bibitem{Unruh}  W. G. Unruh 1981  Phys. Rev. Lett. {\bf 46 } 1351; Unruh 1995
Phys. Rev. D {\bf 51}  (Preprint  gr-qc/9409008)

\bibitem{Jacobson}   T. Jacobson 1991 Phys. Rev. D {\bf 44} 1731 

\bibitem{BLV05}  C. Barcelo, S. Liberati, and M. Visser 2005 Living Rev. Rel. {\bf 8} 12
 (Preprint  gr-qc/0505065)

\bibitem{Foster}  D. Foster 1975 {\it Hydrodynamics Fluctuations, Broken Symmetry, and Correlation
Functions}, (W. A. Benjamin,  Massachusetts)

\bibitem{ZS06}  M. Ortiz de Zarate and J. V. Sengers 2006 {\it Hydrodynamic Fluctuations},
(Elsevier, Amsterdam)


\bibitem{FS08} L.H. Ford and N.F. Svaiter 2008  (Preprint  arXiv:0809.1851)

\bibitem{LL-ST} See, for example, E.M. Lifshitz and  L.P. Pitaevski 1969
{\it Statistical Physics, Part 2}, 2nd ed. (Pergamon, Oxford),
Eq.~(24.10).

\bibitem{FF04} P.O. Fedichev and U. Fischer 2004 Phys. Rev. A {\bf 69}, 033602 

\bibitem{CRC} {\it CRC Handbook of Chemistry and Physics} 2008 D,R. Lide,
  ed., 89th edition (CRC Press, Boca Raton), pp 14-44 and 10-251.

\bibitem{CG66} H.Z. Cummins and R.W. Gammon 1966 J. Chem. Phys. {\bf 44},
2785, Table III.

\bibitem{Jaffe} R. L. Jaffe 2003  (Preprint  arXiv:hep-th/0307014)

\bibitem{MCW06}  K. A. Milton, I. Cavero-Pelaez, and J. Wagner 2006 J. Phys. A {\bf 39},
6543  (Preprint  hep-th/0510236)

\bibitem{SF05}  V. Sopova and  L.H. Ford 2005 Phys. Rev. D {\bf 72}, 105010 
 (Preprint  quant-ph/0504143)

\bibitem{FS98}  L.H. Ford and N.F. Svaiter 1998  Phys. Rev. D {\bf 58}, 065007 
 (Preprint  quant-ph/9804056)

\bibitem{FS08b} L.H. Ford and N.F. Svaiter 2008 manuscript in preparation.

\bibitem{RS03} R. B. Rodrigues and N. F. Svaiter 2003 Physica A {\bf 328}, 466 

\bibitem{CD79} P. Candelas and D. Deutsch 1979  Phys. Rev. D {\bf 20}, 3063 

\bibitem{HK86} T.M. Helliwell and D.A. Konkowski 1986 Phys. Rev. D {\bf 34}, 1918 

\bibitem{FS00}  L.H. Ford and N.F. Svaiter 2000 Phys. Rev. A {\bf 62}, 062105 
 (Preprint  quant-ph/0003129)

\bibitem{FS02}  L.H. Ford and N.F. Svaiter 2002  Phys. Rev. A {\bf 66}, 062106
 (Preprint  quant-ph/0204126)

\bibitem{Vul} V. Vuletic, private communication.

\bibitem{FP06} C.J. Fewster and M.J. Pfenning 2006 J. Math. Phys. {\bf 47}, 082303 



\end{thebibliography}
\end{document}